\documentclass[apl,floatfix,twocolumn]{revtex4}
\usepackage[dvips]{graphicx}
\usepackage{wrapfig}
\usepackage{subfigure}
\usepackage{bm}
\usepackage{color}
\usepackage{amsmath,amssymb}
\usepackage{here}
\usepackage{amsfonts} 
\usepackage{latexsym}

\begin{document}

\title{
Skyrmion-number dependence of spin-transfer torque on magnetic bubbles
}

\author{Yuta Yamane$^1$ and Jairo Sinova$^{1,2}$}
\affiliation{$^1$Institut f\"{u}r Physik, Johannes Gutenberg Universit\"{a}t Mainz,D-55099 Mainz, Germany}
\affiliation{$^2$Institute of Physics ASCR, v.v.i., Cukrovarnicka 10, 162 53 Praha 6, Czech Republic}

\date{\today}

\begin{abstract}
We theoretically study the skyrmion-number dependence of spin-transfer torque acting on magnetic bubbles.
The skymrion number of magnetic bubbles can take any integer value depending on the magnetic profile on its circumference and the size of the bubble. 
We find that the transverse motion of a bubble with respect to the charge current is greatly suppressed as the absolute value of skyrmion number departs from unity, whereas the longitudinal motion is less sensitive.
\end{abstract}

\maketitle

In recent years, attention has been focusing on topologically nontrivial magnetic textures such as magnetic vortices\cite{vortex} and skyrmions\cite{skyrmion}.
They exhibit rich physics stemming from their characteristic structures, which can be advantageous for technological applications\cite{review}.
Another interesting example among such topological textures is magnetic bubbles\cite{textbook};
spot-like closed domains observed in ferromagnetic films with out-of-plane anisotropy, where the magnetization inside the bubble is oriented in the opposite direction to the one outside.
Magnetic bubbles have a potential to play important roles in magnetic memory devices\cite{textbook,skidmore,komineas,moutafis_2007,moutafis_2009,makhfudz,moon,yamane,ogawa,koshibae}.

Vortices, skyrmions and bubbles are quantified by a common topological quantity ${\cal N}_{\rm S}$, the so-called skyrmion number, which is defined by ${\cal N}_{\rm S} = (1/4\pi)\int dxdy \left( {\bm m}\cdot\partial_x{\bm m}\times\partial_y{\bm m} \right)$, where ${\bm m}$ is the classical unit vector in the direction of the local magnetization, and the integral is taken over the film sample.
Whereas a vortex and a skyrmion carry ${\cal N}_{\rm S}=\pm1/2$ and $\pm1$, respectively, for a bubble ${\cal N}_{\rm S}$ can take any integer value depending on the magnetic profile on its circumference and the size of the bubble.
Dynamical response of a bubble to driving forces depends highly on its skyrmion number\cite{moutafis_2009,koshibae};
a tantalizing prospect is that magnetic bubbles with different skyrmion numbers can provide a variety of new functionalities in device applications, which may not be obtained by skyrmions and vortices.

In this work, we theoretically study the ${\cal N}_{\rm S}$-dependence of current-driven bubble motion.
Micromagnetic simulations reveal that the transverse velocity of a bubble with respect to the current is strongly suppressed as $|{\cal N}_{\rm S}|$ departs from unity, while the longitudinal motion is less sensitive.
A collective-coordinate model (CCM), where the steady motion of bubble is assumed, provides good approximate solutions when $|{\cal N}_{\rm S}|=0$, 1, and $|{\cal N}_{\rm S}|\gg1$.

Let us begin by introducing the topological quantities based on which magnetic bubbles can be classified [Fig.~1];
the winding number $S$ counts how many full turns the magnetization on the perimeter of the bubble rotates, and its sign is determined by the sense of rotation.
The polarity $Q$ is defined to take $+1$ when the magnetization inside the bubble points up, and $-1$ when it points down.
The skyrmion number ${\cal N}_{\rm S}$ is given by ${\cal N}_{\rm S} = Q S$.
Below we numerically examine the dependence of current-driven dynamics of a bubble on $Q$ and $S$.
The results of the simulation will be analyzed based on the CCM, where the mathematical expressions for the topological quantities are given.

\begin{figure}[b]
\centering
\includegraphics[width=5.5cm,bb=0 0 425 559]{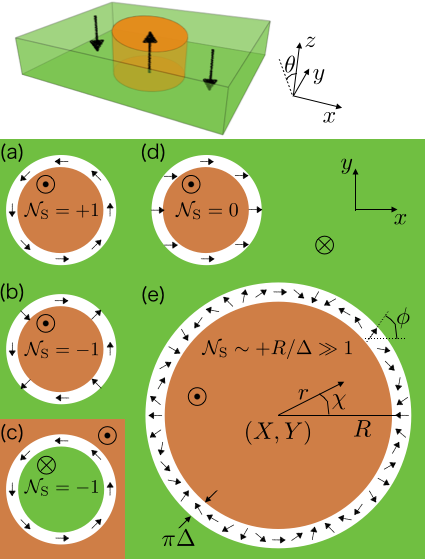}
\caption{ Schematic of magnetic bubbles, where the black arrows indicate the magnetization.
In the bottom figure five topologically different bubbles are shown
 [(a) $S=+1$ and $Q=+1$, (b) $S=-1$ and $Q=+1$, (c) $S=+1$ and $Q=-1$, (d) $S=0$ and $Q=+1$, and (e) $S\simeq + R/\Delta$ and $Q=+1$].
 See the main text for the definitions of the symbols. }
\end{figure}

We assume that the magnetization obeys the Landau-Lifshitz-Gilbert equation;
\begin{equation}
\frac{\partial{\bm m}}{\partial t} =
- \gamma {\bm m} \times {\bm H}_{\rm eff} + \alpha {\bm m} \times \frac{\partial{\bm m}}{\partial t} 
- u \left( 1 - \beta {\bm m} \times \right) \frac{\partial{\bm m}}{\partial x}  ,
\label{llg}\end{equation}
where $\gamma$ is the gyromagnetic ratio, $\alpha$ and $\beta$ are dimensionless parameters, ${\bm H}_{\rm eff}$ is the effective magnetic field due to external, exchange, demagnetizing and anisotropy energies, and $ u = - g \mu_B P j / 2 e M_{\rm S} $ with $g$ the g-factor, $\mu_B$ the Bohr magneton, $M_{\rm S}$ the saturation magnetization, $P$ the spin polarization of the conduction electrons, $e$ the elementary charge, and $j$ the charge current density.
Here the charge current is assumed to flow in the $x$-direction.

\begin{figure}
\centering
\includegraphics[width=5.5cm,bb=0 0 564 748]{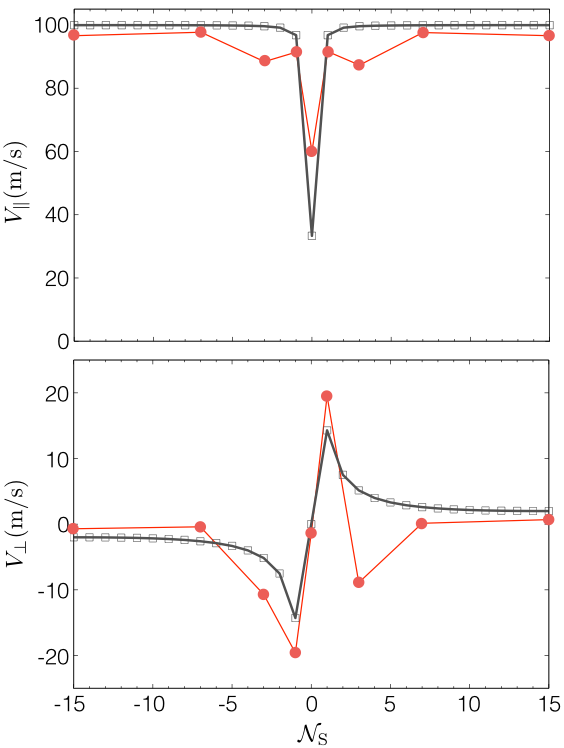}
\caption{ Skyrmion-number dependence of the longitudinal and transverse velocities of bubble in the presence of charge current $u=100$ m/s.
Eqs.~(\ref{eom}) are plotted by the open squares, while the results obtained by micromagnetic simulations are indicated by the red circles.
The solid lines connecting the symbols are guides to the eye.
See the main text for the material parameters used and the combination of $Q$ and $S$ for each ${\cal N}_{\rm S}$.
The two calculations are in good agreement when $|{\cal N}_{\rm S}|=0,1$ and $|{\cal N}_{\rm S}|\sim R/\Delta$.}
\end{figure}

Eq.~(\ref{llg}) is solved by the Object-Oriented Micromagnetic Framework simulator\cite{oommf}, where we divide a square thin film of dimensions $900\times900\times8$ nm$^3$ into $2\times2\times8$ nm$^3$ unit cells, with the material parameters chosen to be typical for Co/Ni;
$\gamma=1.76\times10^{11}$ Hz/T, $M_{\rm S}=6.8\times10^5$ A/m, the uniaxial anisotropy constant $K=4\times10^5$ J/m$^3$, the exchange stiffness $A=10^{-11}$ J/m, $\alpha=0.03$ and $\beta=0.01$.
A magnetic bubble is prepared at the center of the film in equilibrium, and an in-plane current is applied.
We estimate the center-of-mass $(X,Y)$ of a bubble and its radius $R$ by circular fitting to the numerically obtained magnetic profile\cite{moutafis_2009}:
$X=\sum_i (1-m^i_z) x^i f_1(m^i_z) / \sum_i (1-m^i_z)f_1(m^i_z)$, $Y=\sum_i (1-m^i_z) y^i f_1(m^i_z) / \sum_i (1-m^i_z)f_1(m^i_z)$, and $R= \sum_i (1-m^i_z) \sqrt{(x^i-X)^2+(y^i-Y)^2} f_2(m^i_z) / \sum_i (1-m^i_z)f_2(m^i_z)$, where $i$ denotes the unit-cell index, the weighing function $f_1(m_z^i)=1$ when $|m_z^i|<0.99$ but otherwise zero, and similarly $f_2(m_z^i)=1$ when $|m_z^i|<0.1$ but otherwise zero.
The bubble velocity is estimated from the displacement of $(X,Y)$ divided by the time it takes.

In Fig.~2, the results of the simulation is summarized;
the bubble velocity is plotted by the red circles as a function of ${\cal N}_{\rm S}$, where $V_\parallel$ ($V_\perp$) is the longitudinal (perpendicular) velocity with respect to the current.
(For the combination of $Q$ and $S$ employed for each ${\cal N}_{\rm S}$, see the discussion below.)
It is clearly seen that $|V_\perp|$ is greatly suppressed as $|{\cal N}_{\rm S}|$ departs from unity, while $V_\parallel$ is less sensitive to ${\cal N}_{\rm S}$. 
Below we will have a close look at the bubble dynamics at each ${\cal N}_{\rm S}$, and the results for $|{\cal N}_{\rm S}|=0$, $1$, and $15$ will be analyzed by the CCM. 

Fig.~3~(a) shows the equilibrium profile of a bubble with $S=+1$ and $Q=+1$, where the magnetic field $\mu_0 H_z = -2.7$ mT is applied.
$R$ is estimated as $\simeq110$ nm.
A bubble with $S=+1$, $Q=-1$ and the same radius can be obtained exploiting the reversed field $\mu_0 H_z = +2.7$ mT, see Fig.~1~(c) for a schematic.
Fig.~3~(b) are snapshots of the time evolution of the bubble over 2 ns after the current $u = 100$ m/s is turned on.
During the motion, the bubble sustains the circular shape and the magnetization profile shown in Fig.~3~(a).
In Fig.~3~(c), the trajectory of $(X,Y)$ is tracked for three different values of $u$ and $Q=\pm1$ over 2  ns.
The bubbles move with nearly constant velocities after the initial transient regime, and the travel distance is proportional to $|u|$.
The sign change of $Q$ leads to the change in the direction of the transverse motion.
The results for $|{\cal N}_{\rm S}|=1$ shown in Fig.~2 correspond to $S=+1$ and $Q=\pm1$. 
The bubbles with $S=-1$ will be discussed later, where a qualitatively different behaviour than the bubbles with $S=+1$ can be observed.

\begin{figure}
\centering
\includegraphics[width=8cm,bb=0 0 688 690]{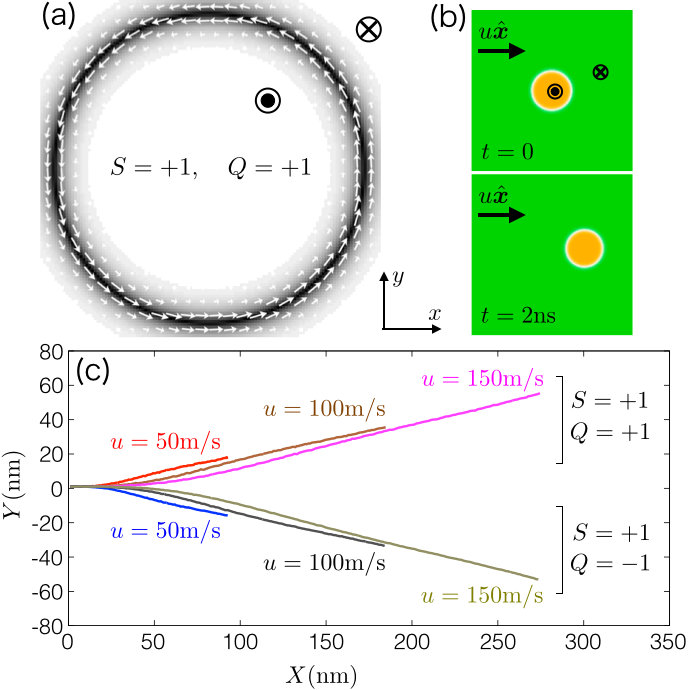}
\caption{ (a) Numerically obtained profile of a bubble with $S=+1$ and $Q=+1$ in equilibrium.
$\mu_0 H_z = -2.7$ mT is applied and $R\simeq110$ nm.
(b) Snapshots of the bubble motion over 2 ns after the current $u=100$ m/s is turned on.
(c) Trajectories of $(X,Y)$ over 2 ns for three different current densities and $Q=\pm1$.
$(X,Y)=(0,0)$ is the initial position at $t=0$, i.e., the center of the thin film.
 }
\end{figure}

Next, let us increase $|{\cal N}_{\rm S}|$ to as large as $15$.
Fig.~4~(a) shows the magnetization configuration at the perimeter of a bubble in equilibrium where $S=+15$, $Q=+1$, $\mu_0 H_z=-9.2$ mT and $R\simeq110$ nm.
Shown in Fig.~4~(b) are the snapshots of the time evolution of this bubble over 2 ns in the presence of current $u=100$ m/s.
The Bloch lines present in the domain wall region are so packed that the dynamics of the magnetization along the perimeter is suppressed enough to sustain the initial state's profile shown in Fig.~4~(a) during the motion.
In Fig.~4~(c), the trajectory of $(X,Y)$ is plotted for the four topologically different bubbles all with $|{\cal N}_{\rm S}|=15$ and the same radius under the current $u=100$ m/s;
the direction of the transverse motion is determined by the sign of ${\cal N}_{\rm S}=QS$.
The linear dependence of the bubble velocity on $|u|$ is indicated in Fig.~4~(d), where the $(X,Y)$-trajectories with $S=+15$ and $Q=\pm1$ are plotted for three different current densities.
The results for $S=+15$ and $Q=\pm1$ are shown in Fig.~2.

\begin{figure}
\centering
\includegraphics[width=8cm,bb=0 0 694 717]{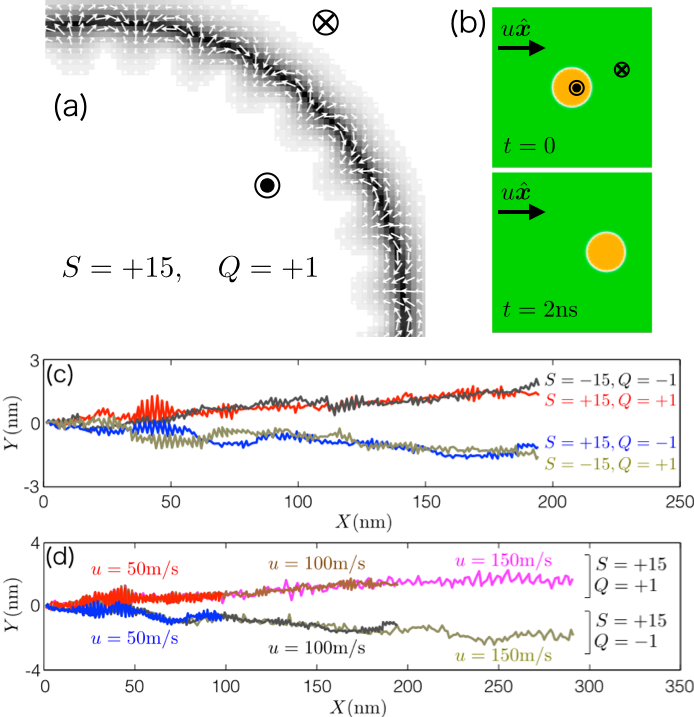}
\caption{ (a) Numerically obtained profile a bubble in equilibrium, where $S=+15$, $Q=+1$, $\mu_0 H_z = - 9.2$ mT, and $R \simeq 110$ nm.
(b) Snapshots of the motion of the bubble over 2 ns in the presence of current $u=100$ m/s.
(c) Trajectories of $(X,Y)$ of the four bubbles with $(S,Q)=(\pm15,\pm1)$ driven by $u=100$ m/s.
(d) Charge current- and $Q$-dependences of the $(X,Y)$-trajectories with $S = +15$.
 }
\end{figure}

A bubble with $Q=+1$ and $S = {\cal N}_{\rm S} = 0$,  i.e., when the magnetic structure is topologically trivial [Fig.~1~(d)], is investigated with magnetic fields $\mu_0 H_z=-11$ mT and $\mu_0 H_x=100$ mT.
The in-plane field is applied to lock the magnetization direction around the circumference in the $x$-direction.
As shown in Fig.~2, $V_\parallel$ is clearly suppressed compared to the other cases, while $V_\perp$ reaches zero.
The small $V_\perp$ is not due to the actual translational motion of the bubble but due to its systematic deformation into an asymmetric ellipse accompanied by the shift of the center-of-mass.

In the three cases discussed above, the bubble shape and the magnetization distribution along the perimeter are rather rigid, motivating us to try to understand the results by a simple analytical model.
Here we assume a perfectly cylindrical bubble with distribution of ${\bm m}=(\sin\theta\cos\phi,\sin\theta\sin\phi,\cos\theta)$ given, as schematically shown in Fig.~1, by\cite{textbook}
\begin{eqnarray}
\theta(r,\chi,z) &=& \pm 2 \tan^{-1}\exp\left[\frac{Q\left(r-R\right)}{\Delta}\right],\label{theta} \\
\phi(r,\chi,z) &=& S\chi + \phi_0,\label{phi}
\end{eqnarray}
where $(r,\chi,z)$ is the cylindrical coordinate measured from the bubble center, $\Delta(\ll R)$ is the domain wall width parameter, $\phi_0$ is a constant.
The topological quantities $Q$ and $S$ are defined by
\begin{equation}
Q=\frac{1}{\pi}\int_0^\infty \frac{\partial\theta}{\partial r}dr =\pm 1,
\end{equation}
and
\begin{equation}
S=\frac{1}{2\pi}\int_{\chi=0}^{2\pi}d\phi = \frac{1}{2\pi}\oint\frac{d\phi}{ds}ds \in {\mathbb Z},
\end{equation}
where $\oint ds$ is the contour integral taken counterclockwise around the circumference of the bubble.
It is straightforward to prove ${\cal N}_{\rm S} = QS$.
There are in general many possible ways of distributing the azimuthal angle $\phi$ along the perimeter, and the linear dependence of $\phi$ on $\chi$ assumed in Eq.~(\ref{phi}) is well satisfied only when $|{\cal N}_{\rm S}|=1$ [Fig.~1~(a)-(c)], ${\cal N}_{\rm S}=0$ with ${\bm m}$ at the perimeter aligned in one direction [Fig.~1~(d)], or $|{\cal N}_{\rm S}|\sim R/\Delta$ where the Bloch lines are packed so closely that the distance between the adjacent Bloch lines is comparable to the domain wall width [Fig.~1~(e)].
When a bubble contains a small number of Bloch lines, the $\phi$-distribution is no longer as simple as Eq.~(\ref{phi}).

Let us employ $(X,Y)$ as the collective coordinate of the bubble dynamics.
Assumed here is the steady motion of the bubble, where the bubble stays rigidly cylindrical with constant radius $R$ during its motion, and the $\phi$-distribution does not change with respect to the comoving coordinates. 
We Integrate Eq.~(\ref{llg}) over the sample volume\cite{tretiakov,clarke} to obtain
\begin{equation}
\left( \begin{array}{c} V_\parallel \\ V_\perp \\ \end{array} \right)
= \frac{u}{G^2 + (\alpha\Gamma)^2}
\left( \begin{array}{c} G^2 + \alpha \beta \Gamma^2 \\ G\Gamma ( \alpha - \beta ) \\
\end{array} \right),
\label{eom}\end{equation}
where ${\bm m} \times {\bm H}_{\rm eff}=0$ has been assumed, and
\begin{eqnarray}
G &=&  4\pi {\cal N}_{\rm S} , \label{g} \\
\Gamma &=&  \iint  \left| \frac{\partial{\bm m}}{\partial x} \right|^2 dxdy \simeq  \frac{2\pi R}{\Delta}   \left( 1 + \frac{{\cal N}_{\rm S}^2\Delta^2}{R^2}  \right) . \label{gamma}
\end{eqnarray}
The equation of motion of the same form with Eq.~(\ref{eom}) has been known for a skyrmion\cite{karin,iwasaki}.
For a bubble, i) owing to the condition $R\gg\Delta$, which is usually not the case for a skyrmion, the analytical expression of $\Gamma$ is accessible as in the second equality of Eq.~(\ref{gamma}),  and ii) $|{\cal N}_{\rm S}|$ is not restricted to $1$, leading to the strong ${\cal N}_{\rm S}$-dependence of $ ( V_\parallel, V_\perp ) $ that enriches the bubble dynamics as already seen.
Eq.~(\ref{eom}) is compared to the numerical results in Fig.~2 by the open symbols.
The parameters are chosen to be consistent with the simulation at each ${\cal N}_{\rm S}$.
The two calculations agree well at $|{\cal N}_{\rm S}|=0$, 1, and 15.

Lastly, we touch upon a couple of cases where the CCM is not a good approximation.
A bubble with $S=-1$ [Fig.~1 (b)] inevitably produces magnetic charges on its perimeter and thus is energetically unfavourable.
This fact leads to relatively large shape distortion of the bubble during its motion, losing the legitimacy of using the CCM.
By the micromagnetic simulation (not shown) with $S=-1$, $Q=+1$, $\mu_0 H_z = - 2.2$ mT, $R \simeq 110$ nm and $u=100$ m/s, we observed $V_\parallel\simeq 98.4$ m/s and $V_\perp\simeq - 2.3$ m/s;
whereas $V_\parallel$ agrees well with Eq.~(\ref{eom}), $|V_\perp|$ is about of an order smaller than the prediction by the CCM.
Shown in Fig.~5 is a case with ${\cal N}_{\rm S} = + 3$.
The two pairs of Bloch lines move along the circumference in the presence of charge current.
The numerical results with $S=+3$, $Q=\pm1$, and the Bloch-line distribution shown in Fig.~5 are compared with Eq.~(\ref{eom}) in Fig.~2;
the CCM completely fails to predict $V_\perp$.
We also observed that the bubble dynamics depends highly on the initial locations of the Bloch lines (not shown).
The signal of the restoration of agreement between the CCM and the simulation is seen when $|{\cal N}_{\rm S}|$ is increased to 7.
We leave more systematic and complete investigations to future work.

\begin{figure}
\centering
\includegraphics[width=5.5cm,bb=0 0 434 529]{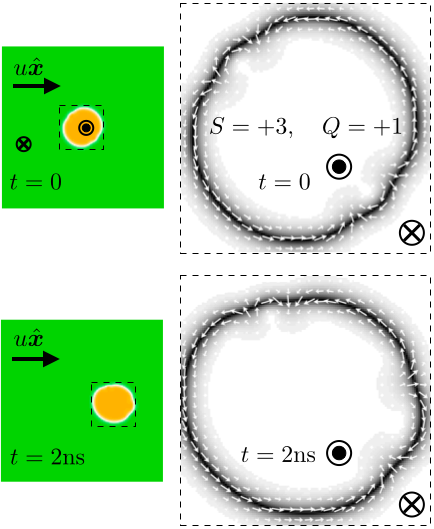}
\caption{ Snapshots of the time evolution of the bubble with $S=+3$ and $Q=+1$ over 2 ns in the presence of current $u=100$ m/s and the magnetic field $H_z = - 3.1$ mT.}
\end{figure}

In conclusion, we presented analytical and numerical studies on the current-driven bubble motion.
We found that the transverse motion of the bubble with respect to the current is greatly suppressed as the bubble's skyrmion number departs from unity. 
Our findings suggest the possibility to manipulate the dynamics of bubbles by their skyrmion number, which would lead to implementation of magnetic bubbles in wider range of applications.

We are grateful to Peng Yan and Jun'ichi Ieda for their valuable comments on this work.
This research was supported by Research Fellowship for Young Scientists from t, Alexander von Humboldt Foundation, the Ministry of Education of the Czech Republic (Grant No. LM2011026) and the Grant Agency of the Czech Republic (Grant No. 14-37427).

\end{document}